\documentclass{aa}

\usepackage{graphicx}
\usepackage{lipsum}
\usepackage{natbib}
\usepackage{txfonts}
\begin{document}

   \title{Magnetic activity of the young solar analogue V1358 Ori}

   \author{L. Kriskovics
          \inst{1}
          \and
          Zs. K\H{o}v\'ari
          \inst{1}
          \and
          K. Vida
          \inst{1}
          \and
          K. Ol\'ah
          \inst{1}
          \and
          T. A. Carroll
          \inst{2}
          \and
          T. Granzer
          \inst{2}
          }

   \institute{Konkoly Observatory, Research Center for Astronomy and Earth Sciences, Budapest, Hungary\\
              \email{kriskovics.levente@csfk.mta.hu}
         \and
             Leibniz Institute for Astrophysics (AIP), Potsdam, Germany}

   \date{Received ...; accepted ...}

  \abstract
   {Young, fast rotating single stars can show dramatically different magnetic signatures and levels of magnetic activity as compared with the Sun. 
   While losing angular momentum due to magnetic breaking and mass loss through stellar winds, the stars gradually spin down resulting in decreasing levels of activity.
   Studying magnetic activity on such solar analogues plays a key role in understanding the evolution of solar-like stars and allows a glimpse into the past of the Sun as well.}
   {In order to widen our knowledge of the magnetic evolution of the Sun and solar-like stars, magnetic activity of the young solar analogue V1358\,Ori is investigated.}
   {Fourier analysis of long-term photometric data is used to derive rotational period and activity cycle length, while spectral synthesis is applied on 
   high resolution spectroscopic data in order to derive precise astrophysical parameters.
   Doppler imaging is performed to recover surface temperature maps for two subsequent intervals. 
   Cross-correlation of the consecutive Doppler maps is used to derive surface differential rotation. The rotational modulation of the chromospheric activity indicators is also investigated.}
   {An activity cycle of $\approx$1600 days is detected for V1358 Ori. 
   Doppler imaging revealed a surface temperature distribution dominated by a large polar cap with a few weaker features around the equator. This spot configuration is similar to other maps of young solar analogues from the literature, and supports recent model predictions.
   We detected solar-like surface differential rotation with a surface shear parameter of $\alpha=0.016\pm0.010$ which fits pretty well to our recently proposed empirical relation between rotation and differential rotation. The chromospheric activity indicators showed a rotational modulation.}
   {}

   \keywords{stars: activity --
                stars: imaging  --
                starspots --
                stars: individual: V1358 Ori
               }

   \maketitle
%
%________________________________________________________________

\section{Introduction}
   \begin{figure*}[t!!]
   \resizebox{\hsize}{!}
            {\includegraphics[clip]{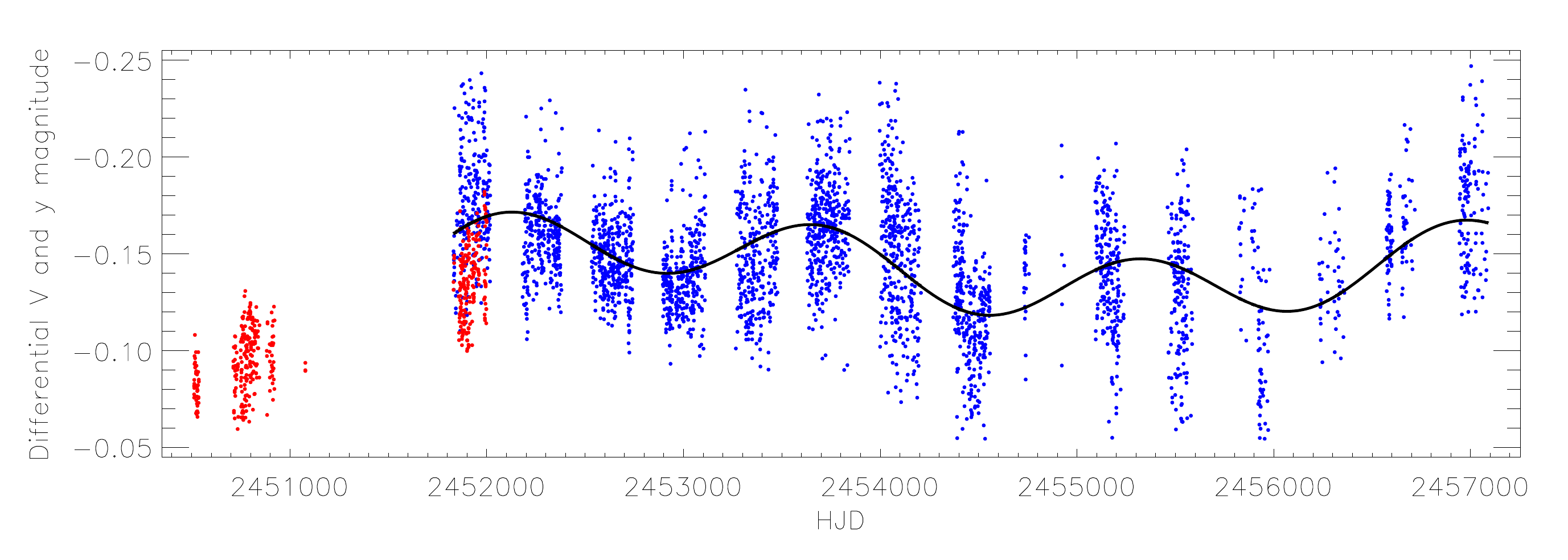}}
      \caption{Strömgren $y$ (red) and Johnson $V$ (blue) differential photometry of V1358\,Ori. The overplotted curve is the combination of the two long-term cycles. For the long-term, fit we used the Johnson $V$ data only, and the Stromgren $y$ data were left out because of the ~0.025 mag shift which is probably due to the different transmission characteristics the two passbands. See text for details.}
         \label{fig_phot}
   \end{figure*}

Studying magnetic activity -- indicators of the state of the underlying magnetic dynamo --  plays a key role in understanding the evolution of solar-like stars along the main sequence, since the magnetic dynamo strongly affects not just the stellar structure \citep{berd_review}, but the spin--down of the star due to magnetic breaking (e.g. \citealt{barnes_rot}), angular momentum loss through stellar winds \citep{macgregor_wind}, etc. 

Strong magnetic activity could  also affect orbiting planetary systems through strong stellar winds or high-energy electromagnetic or particle radiation, which may ultimately erode planetary atmospheres as well (e.g. \citealt{vida_trappist}).

Solar magnetic fields are generated by an $\alpha\Omega$-type dynamo \citep{parker_dyn}. Here, the poloidal field is wind up and amplified by the differential rotation creating the toroidal field ($\Omega$ effect), while the $\alpha$ effect creates small-scale poloidal fields from the toroidal field via  rotationally induced convective turbulence, and turbulent diffusion builds up the large scale poloidal field by reconnection of the small-scale field.

In rapidly rotating late-type (G--K) dwarfs, the $\Omega$-effect can be suppressed, resulting in an $\alpha^2\Omega$ type dynamo (see \citealt{ossen_dyn}, and references therein, or e.g. \citealt{kovari_lqhya} for observational evidence).  However, as these stars evolve, they spin down mostly due to magnetic breaking (\citealt{skumanich1}, \citealt{barnes_rot}), thus their dynamo shift from the $\alpha^2\Omega$ domain to the $\alpha\Omega$ \citep{ossen_dyn}.

In order to properly understand the magnetic evolution of the Sun and solar-like stars along the main sequence (and its effect to their vicinity), it is imperative to study the magnetic activity of young solar-type stars. 

V1358 Ori (HD 43989, HIP 30030) was originally identified as an active star and potential Doppler imaging target and classified as G0IV by \cite{strassmeier_kpnovienna}. Later it was reclassified as an F9 dwarf by \cite{montes_v1358ori}, which was confirmed by \cite{mcdonald_v1358}. \cite{vican_v1358ori} estimated the effective temperature to be $T_{\mathrm{eff}}=6100\,\mathrm{K}$. They also reported infrared excess based on WISE data and suggested the presence of a debris disk around the star. \cite{zuckerman_col} identifies V1358\,Ori as a member of the Columba association, and thus, a very young ($\approx 30\,\mathrm{Myr}$) solar analogue.

\cite{hackman1} carried out a Zeeman--Doppler analysis of V1358\,Ori. They reported a strong toroidal magnetic field component on the Stokes V maps, and prominent polar features on the brightness map, as well as some weaker, lower latitude spots, however their phase-coverage was poor.

In this paper, we carry out a photometric and spectroscopic analysis of the young solar-analogue V1358\,Ori, based on a 18 years long, homogeneous photometric data set and Doppler imaging applied on high-resolution spectra covering two rotations. We also derive surface differential rotation from the consecutive Doppler maps and compare the results to similar young active solar analogues.

%__________________________________________________________________

\section{Observations}
Photometric data of Strömgren $by$ and Johnson-Cousins $VI_{\mathrm{C}}$ data were gathered with \emph{Wolgang} and \emph{Amadeus}, the 0.75-m automatic photoelectric telescopes operated by the Leibniz Institute for Astrophysics Potsdam, and located at Fairborn Observatory, Arizona \citep{strassmeierapt} between 4 Mar 1997 and 8 Mar 2015. All differential measurements were taken respect to the comparison star HD\,44517. Between 4 Mar 1997 and 16 Apr 1997, the check star was HD\,44019. After that, HD\,45215 was used. For details on the data reduction, see \cite{strassmeierapt} and \cite{granzerapt}. Photometric $Vy$ data are plotted in Fig. \ref{fig_phot}. We note that there is a ~0.025 mag shift between the Stromgren $y$ data of the first two seasons and the rest of the observations, which is probably due to the different transmission characteristics of the two passbands.

Spectroscopic observations were gathered via OPTICON with the NARVAL high-resolution echelle spectropolarimeter mounted on the 2-m Bernard Lyot Telescope of Observatoire Midi-Pyr\'en\'ees at Pic du Midi, France between 09-20 Dec 2013. A peak resolution of $R=80000$ was reached in spectroscopic object mode. The exposure time was $t_{\mathrm{exp}}=1200\,\mathrm{s}$, yielding a typical signal-to-noise ratio of $\approx300$ (hereafter S/N) at 6400\,\AA{} (see Table \ref{table_spec} for details).

\begin{table}
\caption{Spectroscopic observing log for V1358 Ori. S1 and S2 indicate the two subsets used for Doppler imaging,HJD$-$2456000, $\phi$ and S/N are the reduced Heliocentric Julian dates, rotational phases and signal-to-noises, respectively.}             
\label{table_spec}      
\centering                         
\begin{tabular}{c c c || c c c c}        
\hline\hline
& S1  & & & S2  & \\
\hline\hline
HJD-2456000 & $\phi$ & S/N & HJD-2456000 & $\phi$ & S/N \\
\hline
636.4313 & 0.848 & 314 &  641.4184 & 0.523 &242 \\                 
636.6900 & 0.039 & 361 &  641.6738 & 0.711 &289 \\
637.4258 & 0.581 & 332 &  642.4252 & 0.265 &299 \\
637.6757 & 0.765 & 321 &  643.3967 & 0.981 &335 \\
638.6755 & 0.502 & 342 &  647.3944 & 0.926 &332 \\
639.4232 & 0.053 & 312 &  647.6529 & 0.117 &347 \\
639.6681 & 0.233 & 317 &   & & \\

\hline
\end{tabular}
\end{table}

To phase the observations, the following ephemeris was used:
\begin{equation}
    \mathrm{HJD}=2\,449\,681.5+1.3571\times E
\end{equation}
$T_0$ was adopted from \cite{hackman1}. For details on the rotational period, see Sect. \ref{subsect_phot}.

Spectroscopic data reduction were carried out with the standard NARVAL data reduction pipeline. ThAr arc-lamps were used for wavelength calibration. An additional continuum fit and normalization were applied in order to avoid erroneous continuum fits during the spectral synthesis and Doppler inversion.

\section{Photometric analysis}\label{subsect_phot}

The period analysis were carried out on the Johson $V$ data. Strömgren $y$ data were excluded as it only consist of 458 points and there is a long gap at the beginning of the time series (see Fig. \ref{fig_phot}).

For photometric period determination, we used  MuFrAn\footnote{\url{https://konkoly.hu/staff/kollath/mufran.html}}, a code for frequency analysis based on Fourier-transformation (\citealt{kollath96}, \citealt{csubry_mufran}). We accepted the peak at the  cycle-per-day value $\mathrm{c/d}=0.7368$ yielding $P_{\mathrm{rot}}=1.3571\,\mathrm{d}$ as the rotational period, which is close to the period derived by \cite{cutispoto} ($P_{\mathrm{rot}}=1.16\,\mathrm{d}$). The reasoning behind this is the following. We gradually pre-whitened the Fourier-spectra with the 
two peaks corresponding to the two long-term changes ($\approx\!1600$ and $\approx\!5200$ days, see Fig. \ref{fig_phot_fourier}), the suspected rotational period and the "forest" of peaks around that value, and a signal of one day caused by the binning of the data (two or three consecutive measurements were taken in $V$ and $I$ each day). The resulting Fourier-spectrum contained no signal which could reasonably be considered real within the precision of the photometry, and since the only period is the 1.3571 d which cannot be attributed to artificial origin or a long-term cyclic behaviour, we accepted it as the rotational period. Moreover, only $P_{\mathrm{rot}}=1.3571\,\mathrm{d}$ yields a properly phased light curve. The result is very close to the average of the seasonal periods ($P_{\mathrm{rot}}=1.3618\,\mathrm{d}$) and is also consistent with the measured $v\!\sin\!i$ value (see Sect. \ref{sect_fund}). 
\begin{figure}[t!!]
\resizebox{\hsize}{!}
            {\includegraphics[clip]{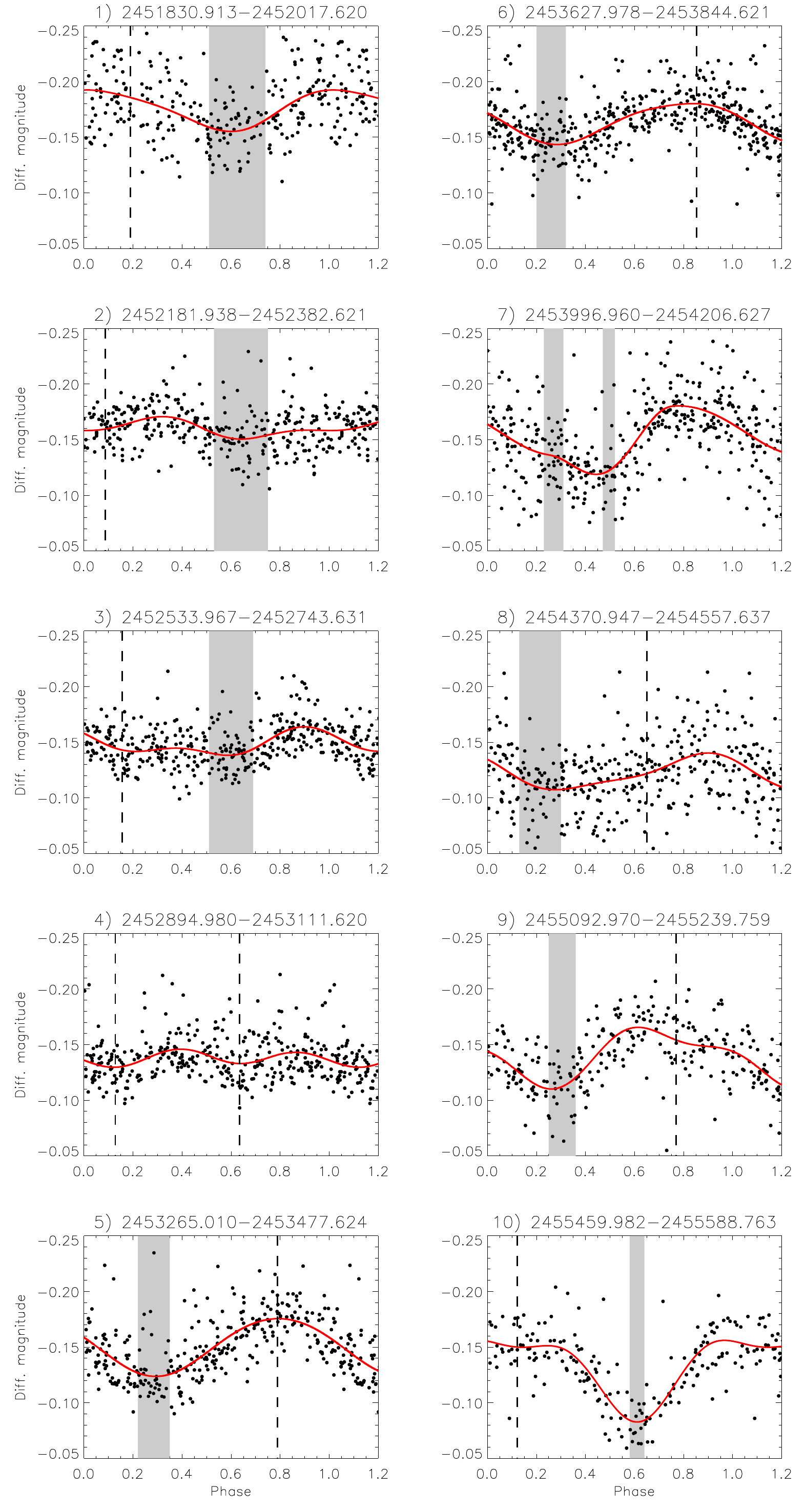}}
      \caption{Seasonal light curves of V1358 Ori phased with $P_{\mathrm{rot}}=1.3571\,\mathrm{d}$. The time periods are denoted over the plots in JD along with the number of the seasons.   Solid (red) lines show a two-spot spot model fitted to the light curves, while the gray zones and dashed lines show the longitudinal positions of the fitted spots. Seasons where the low number of data points made visual inspection meaningless were omitted. 
              }
         \label{fig_phot_all}
\end{figure}

Fig.~\ref{fig_phot_all} shows the seasonal phased $V$ light curves to demon\-strate the robustness of the derived photometric period. Solid lines indicate a two-spot analytic spot model fitted to each season using SML \citep{ribarik_sml}. Shaded zones show the position of the dominant spot, the width of the zone denotes the error of the spot longitude. Vertical dashed lines indicate the weaker spot (on the plot of the fourth season, both active longitudes are weak, while in case of the seventh season, it is hard to decide which is the dominant longitude). Seasons where the low number of points made the inspection impossible were omitted from the plot: two between $\mathrm{HJD}=2454557$ and 2455092, one between 2455239 and 2455459, and two after 2455588. On subplots 4) and 5) (intervals 2452895.0--2453111.6 and 2453265.0--2453477.6), the active longitudes seem to shift from $\approx$0.1 and  $\approx$0.6 to  $\approx$0.3 and  $\approx$0.8, while on subplot 10) (2455460.0--2455588.8) they appear to be in the previous positions again. This could indicate the presence of a flip-flop like phenomenon \citep{jetsu_flipflop}, with a time scale of  $\approx$6 years. 

By visual inspection of the complete 14 years long $V$ light curve, a long-term cyclic behaviour can be suspected as well, which was confirmed by the Fourier-analysis, yielding $P_{\mathrm{cyc}}\approx1600\,\mathrm{d}$ (see the first Fourier-spectrum plot in Fig. \ref{fig_phot_fourier}).The sum of this cycle and the other long-term change from the Fourier-analysis ($\approx 5200\,\mathrm{d}$) is overplotted on the complete $V$ light curve in Fig. \ref{fig_phot}. There is a ~0.025 mag systematic shift between the $V$ and $y$ data, which is probably due to the different transmission characteristics of the two bands. Therefore we decided not to include the Strömgren data of the first two seasons in our long-term analysis.

For further discussion on the spot configuration, suspected flip-flop and the activity cycle, see Sect.
 \ref{sect_disc}.

\begin{figure}[t!!]
\resizebox{\hsize}{!}
            {\includegraphics[clip]{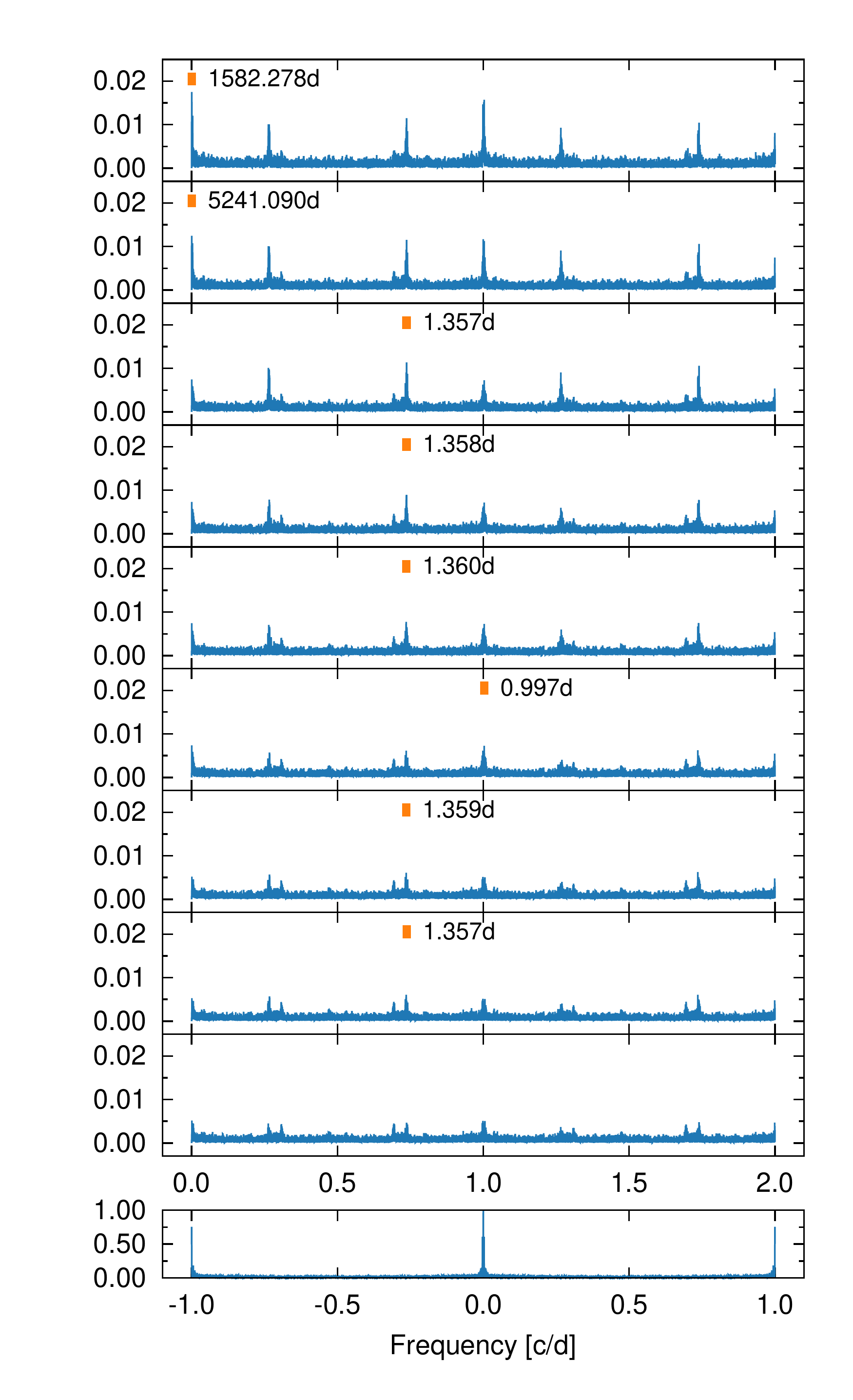}}
      \caption{Upper panels: Fourier spectra of the $V$ light curve of V1358 Ori.  Orange rectangles denote the strongest peaks of the subsequent steps. Consecutive plots show pre-whitened spectra with the periods shown. Bottom panel: Fourier window-function. See Sect. \ref{subsect_phot} for details. 
              }
         \label{fig_phot_fourier}
\end{figure}

\section{Fundamental parameters}\label{sect_fund}

\subsection{Spectroscopic analysis}

Precise astrophysical parameters are fundamental for Doppler inversion, therefore we carried out a detailed spectroscopic analysis based on spectral synthesis using the code SME \citep{piskunov_sme}. During the synthesis, MARCS models were used \citep{gustafsson_marcs}. Atomic line parameters were taken from the VALD database \citep{kupka_vald}. Macroturbulence was estimated using the following equation \citep{valenti_macro}:

\begin{equation}
    v_{\mathrm{mac}}=\Bigg(3.98-\frac{T_{\mathrm{eff}}-5770 \mathrm{K}}{650 \mathrm{K}}\Bigg)\,\mathrm{km\,s}^{-1}
\end{equation}

Astrophysical parameters were determined using the following methodology:
   
\begin{enumerate}
    \item Determination of $v\sin i$ using initial astrophysical parameters taken from \cite{montes_v1358ori} and assuming solar abundances.
    \item Microturbulence ($\xi$) fit with same astrophysical parameters and $v\sin i$ from step 1 using lines with $\log gf<-2.5$, since weak lines are more sensitive to the change of microturbulence.
    \item Refitting $v\sin i$ and $\xi$ simultaneously to check robustness.
    \item Fitting $T_{\mathrm{eff}}$ using solar metallicity and line broadening parameters from step 3.
    \item Fitting metallicity using the effective temperature value from step 4.
    \item Fitting $\log g$ using stronger lines ($\log gf > 0$).
    \item Refitting $T_{\mathrm{eff}}$, $\log g$ and metallicity simultaneously to check robustness.
    \item Refitting $v\sin{i}$.
\end{enumerate}

Lithium abundance fit was carried out using NLTE departure coefficients \citep{piskunov_sme}. The fit yielded $A(\mathrm{Li})=2.27\pm0.05$. An example of the fit can be seen in Fig \ref{fig_li}.
The astrophysical parameters are summarized in Table \ref{table_pars}.

\begin{table}[b!!] 
\caption{Fundamental astrophysical parameters of V1358 Ori.}
\label{table_pars}
\centering 
\begin{tabular}{c|c}
\hline\hline
$T_{\mathrm{eff}}$ & $6040\pm25\,\mathrm{K}$\\
$\log{g}$  & $4.44\pm0.04$ \\
$[\mathrm{Fe}/\mathrm{H}]$& $0.04\pm0.02$ \\
$v_{\mathrm{mic}}$ & $3.0\pm0.5\,\mathrm{km\,s}^{-1}$ \\
$v_{\mathrm{mac}}$ (computed )& $ 3.6\,\mathrm{km\,s}^{-1} $ \\
$v\sin{i}$ & $38\pm1\,\mathrm{km\,s}^{-1}$\\
Distance & $52.0\pm1.3\,\mathrm{pc}$ \\
$M_{\mathrm{bol}}$&$4.23^{\mathrm{m}} {}^{+0.06}_{-0.05}$\\
$L/L_{\odot}$ & $1.62^{+0.09}_{-0.07}$ \\
$R/R_{\odot}$ & $1.17 \pm 0.03$ \\
Inclination & $60\pm10 \, {}^{\circ}$ \\
$P_{\mathrm{rot}}$ & $1.3571\,\mathrm{d}$ \\
NLTE Li abundance & $2.27\pm0.05$ \\
\hline\hline
\end{tabular}
\end{table}

\subsection{Distance, radius, inclination}
The \emph{Gaia} DR2 parallax of $\pi=19.22\pm0.05 \, \mathrm{mas}$ (\citealt{gaia,gaia_dr2}) gives a distance of $d=52\pm 1.3 \, \mathrm{pc}$. This distance using the average $V$ brightness of the star yields a bolometric magnitude of 
$M_{\mathrm{bol}}=4.23^{+0.06{\mathrm{m}}}_{-0.05}$ (with extinction from \citealt{schlafly_redd} and bolometric correction from \citealt{flower_bc} taken into account). This results in a luminosity of 
$L/L_{\odot}=1.62^{+0.09}_{-0.07}$, which is in good agreement with the value from \emph{Gaia} DR2 ($L/L_{\odot}=1.64\pm0.015$, \citealt{gaia_dr2}), and thus a radius of $R/R_{\odot}=1.17\pm0.03$. The radius with the photometric period and the $v\sin i=38\pm 1 \, \mathrm{km\,s^{-1}}$ from the spectral synthesis yields an inclination of $i=60\pm 10^{\circ}$.

\subsection{Age}
The lithium abundance from the spectral synthesis using the empirical correlation between age and abundance from \cite{carlos_liage} would yield $t  \approx 0.75\pm0.5\,\mathrm{Gyr}$. However, it was pointed out by e.g. \cite{balac_li} or \cite{eggenberger_lirot} that in case of fast-rotating stars, lithium abundance is not always a suitable age indicator. Li depletion is also dependent on the spectral type: on an M dwarf, Li can be depleted in ~ 10 Myrs, while on a hotter star, this rate is much lower.

According to e.g. \cite{zuckerman_col}, V1358 Ori is a member of the Columba association, which has an age of $\approx 30\,\mathrm{Myr}$. This is in good agreement with the computed gyrochronological age ($t_{\mathrm{gyro}}=23\pm7\,\mathrm{Myr}$, using Eq. 5.3 in \citealt{barnes_gyro}). Thus, most likely, V1358 Ori is a very young solar analogue.

\begin{figure*}[t!!]
            
            \vspace{2.0cm}
            \Large{S1}
            
            \vspace{-2.3cm}
            \hspace{1.5cm}\includegraphics[width=0.9\textwidth]{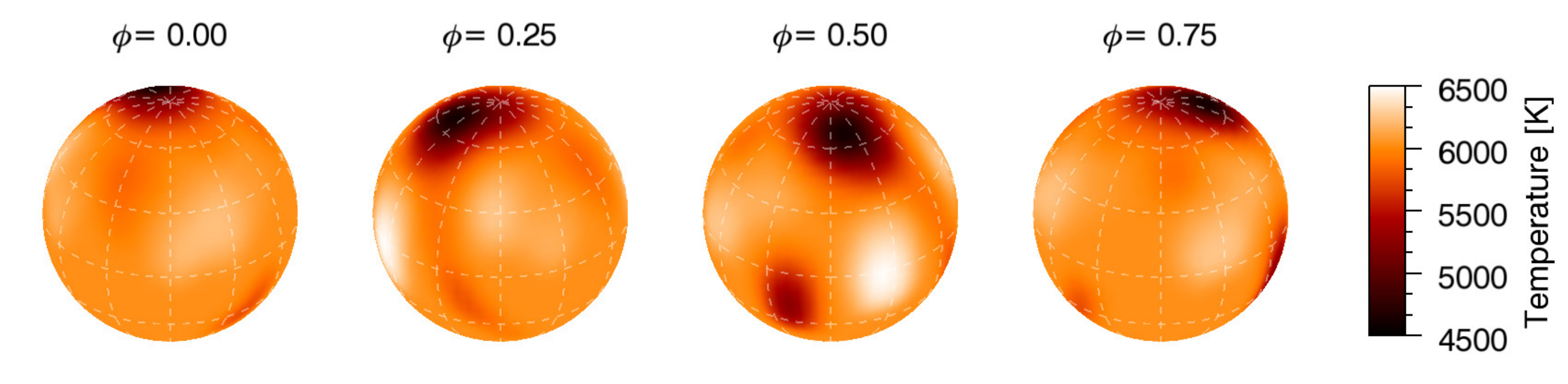}
    
            \vspace{2.0cm}
            \Large{S2}
            \vspace{-2.3cm}
            
            \hspace{1.5cm}\includegraphics[width=0.9\textwidth]{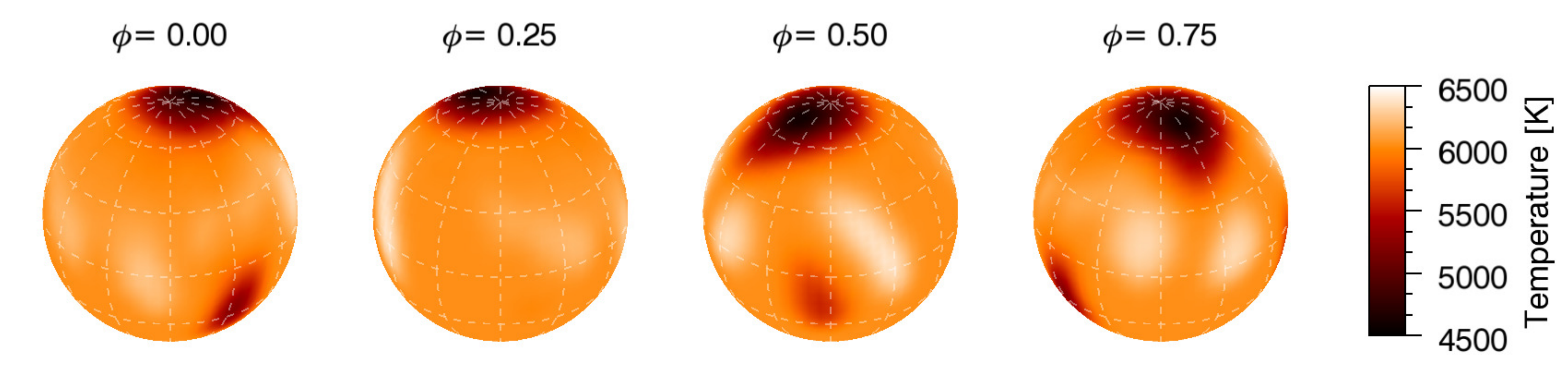}
            
      \caption{The two consequent Doppler images of V1358 Ori plotted in four different rotational phases. The corresponding average HJDs for the maps are 2456638.1 and 2456653.9 for S1 and S2 respectively.
              }
         \label{fig_di}
\end{figure*}

\begin{figure}[t!!!]
    \begin{minipage}[b]{0.45\linewidth}
        \Large{S1}
        \centering
        \includegraphics[width=\textwidth]{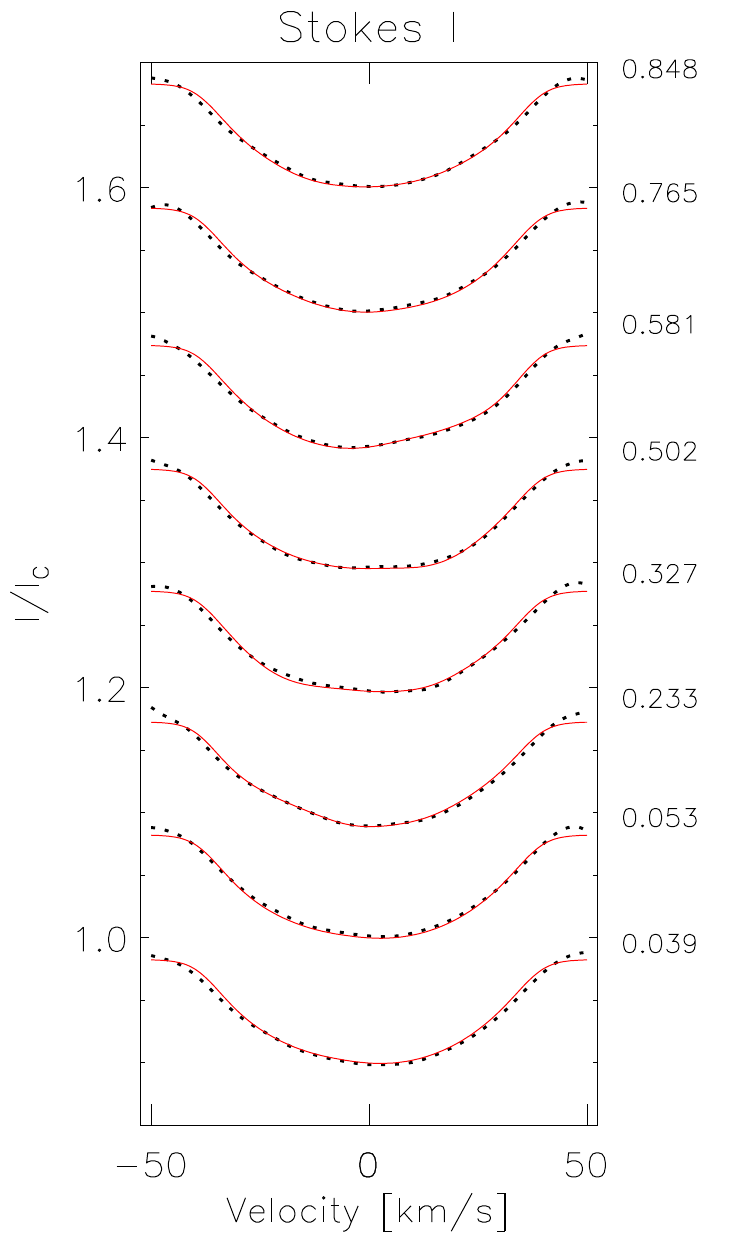}
    \end{minipage}
    \begin{minipage}[b]{0.45\linewidth}
        \Large{S2}
        \hspace{0.5cm}
        \centering
        \includegraphics[width=\textwidth]{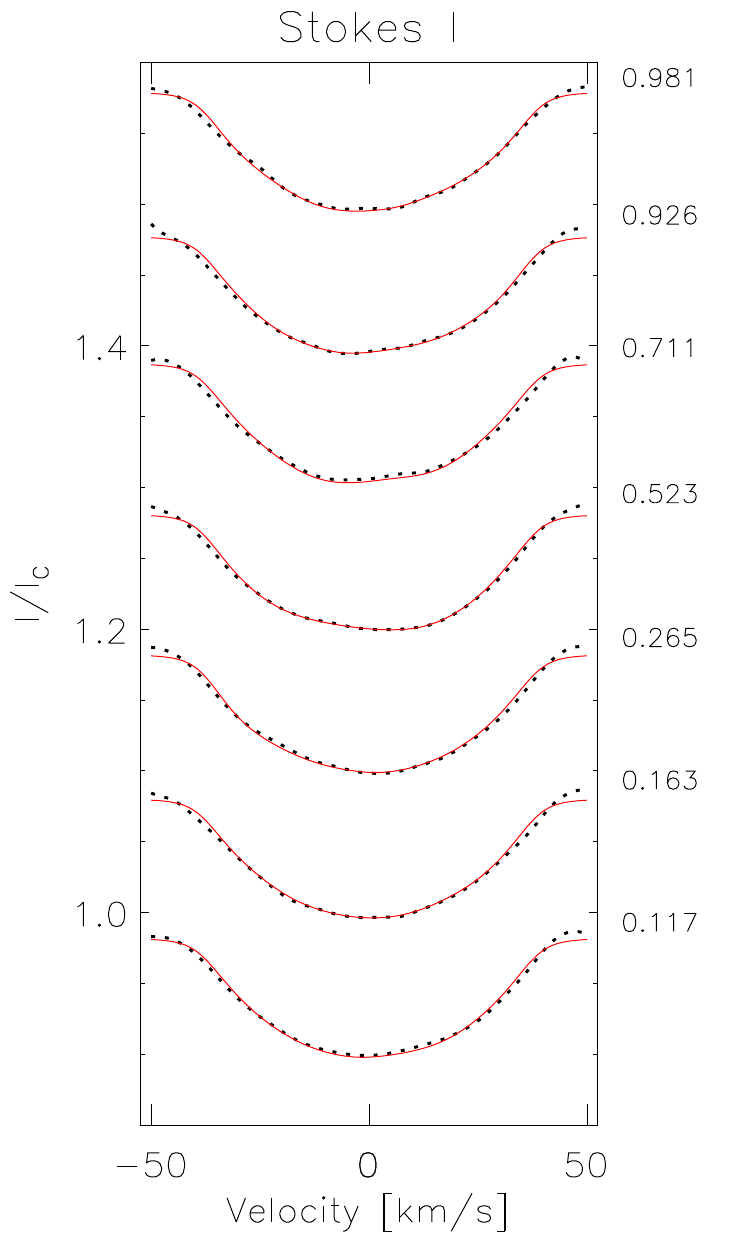}
    \end{minipage}
      \caption{Observed averaged line profiles and final fits for the two subsets. The dotted (black) lines represent the observations, while the solid (red) lines are the fits from the Doppler imaging. . Rotational phases are denoted on the right side of the plots. 
              }
         \label{fig_prof}
\end{figure}

\section{Doppler imaging}

\subsection{The next generation code iMap}
Our Doppler imaging code \emph{iMap} \citep{carroll_imap} carries out multi-line Doppler inversion on a list of photospheric lines between 5000--6750\,\AA. We included 40 virtually non-blended absorption lines with suitable line-depth, temperature sensitivity and well defined continuum. The stellar surface is divided into $5^{\circ}\times 5^{\circ}$ segments. For each local line profile, the code utilizes a full radiative solver \citep{carroll_solver}. Then the local line profiles are disk integrated, and the individually modeled,  disk-integrated lines are averaged. Atomic line data are taken from VALD \citep{kupka_vald}. Model atmospheres are taken from \cite{castelli_mod} and are interpolated for the necessary temperature, gravity or metallicity values. Due to the high computational capacity requirements, LTE radiative transfer is used instead of spherical model atmospheres, but imperfections in the fitted line shapes are well compensated by the multi-line approach.  Additional input parameters are micro- and macroturbulence, and $v\!\sin\!i$.

For the surface reconstruction, an iterative regularization method based on a Landweber algorithm is used \citep{carroll_imap}, meaning no additional constraints are imposed in the image domain. According to our tests (Appendix A in \citealt{carroll_imap}), the iterative regularization proved to be effective and inversions based on the same datasets always converged to the same image solution.

\subsection{Surface reconstructions}\label{sect_di}
The available 15 spectra are divided into two subsets. The corresponding time intervals are 2456636.43--2456639.66 and 2456641.42--2456647.65. The first subset consists of 8 spectra and covers 2.4 rotations, while the other 7 spectra of the second subset covers 4.6 rotations. The phase coverages are not completely uniform, nevertheless both subsets are suitable for Doppler imaging.

The resulting two Doppler reconstructions (henceforth S1 and S2) for V1358\,Ori are plotted in Fig.~\ref{fig_di}. The average profiles are plotted along with the final profile fits (thick black and thin red lines, respectively) in Fig \ref{fig_prof}. 

\begin{figure*}[t!!]
    \resizebox{\hsize}{!}
            {\includegraphics[clip]{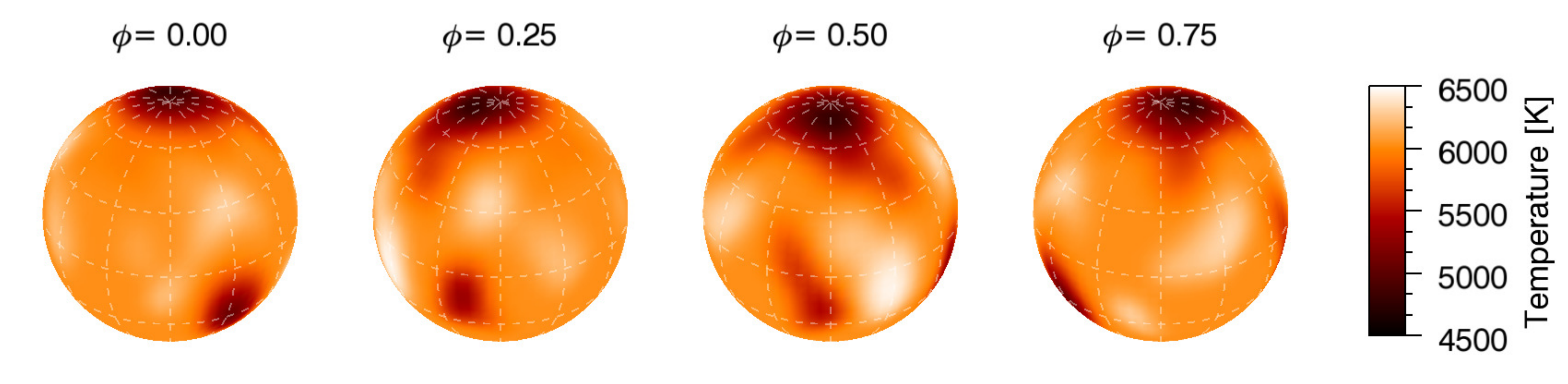}}
      \caption{Average Doppler image of V1358\,Ori derived using all of the spectra plotted in four rotational phases.
              }
         \label{fig_diall}
\end{figure*}

The overall characteristics of the two individual Doppler reconstructions are quite similar. This is supported by the average Doppler image using all the available spectra, see Fig.~\ref{fig_diall}. The resulting average map is pretty similar to the two individual images in Fig.~\ref{fig_di}, indicating only minor changes in the spot configuration. 

Doppler imaging reveals a strong polar cap, as well as both cool and hot features at lower latitudes, down to the equator. Spot temperatures range from $\approx\!4500\,\mathrm{K}$ to $\approx\!6400\,\mathrm{K}$, this latter is $\approx\!350\,\mathrm{K}$ higher than the temperature of the quiet photosphere. The contrast of the coolest features of $\approx\!1500\,\mathrm{K}$ relative to the unspotted photosphere is around the typical values for a late F, early G dwarf \citep{berd_review}.

The shape of the polar feature changes somewhat from S1 to S2: while on S1, the spot features an appendage reaching~$\approx\!45^\circ$ latitude around 0.4 phase, another eccentricity is seen at $\approx\!0.7$ on S2. There is a fainter cool feature around 30$^\circ$ latitude and 0.9 phase, which almost completely disappears on S2, however the sub-equatorial spot around 0.2 phase becomes more prominent. The other cool low latitude features at $\approx\!0.6$ phase fades somewhat, and it is also displaced to $\approx\!0.55$.

There are bright features present on both maps. There is a larger one at $\approx\!0.4$ phase which becomes somewhat less prominent and of different shape on the second image. There are also other much weaker equatorial bright spots on both maps, with slightly different shapes. 

Hot features are often considered to be of artificial origin, mostly caused by insufficient phase coverage, as previous tests (e.g. \citealt{lindborg1}) have pointed out. However, decreasing the phase coverage usually introduces both cool and hot features on roughly the same longitude (Fig. 3 in \citealt{lindborg1}). Also, the two Doppler maps both show similar hot features at the same positions, and are based on completely independent datasets with different phase coverages. These features can also be seen on the average map derived from all of the spectra, where the phase coverage is inherently better, roughly at the same positions. The shape of the chromosperic activity indicator curves might also support the conclusion that the bright spots are real (see Sect. \ref{sect_chrom}). Thus, we conclude that it is more likely that these hot spots are indeed real features.

\subsection{Differential rotation}
Longitudinal spot displacements from the first series compared to the second can be used as a tracer of surface differential rotation. Visual inspection of the two subsequent Doppler maps may indicate such rearrangements (see Sect. \ref{sect_di}): the longitudinal displacement of the subequatorial cool spot around 0.4 phase or displacement and change of shape of the bright feature at $\approx\!0.6$ phase can be interpreted as such.

Surface differential rotation can be measured by longitudinally cross-correlating consecutive Doppler images \citep{donati_ccf}, and fitting the latitudinal correlation peaks by an assumed quadratic rotational law (see e.g. \citealt{kovari_zetand}):  

\begin{equation}
    \Omega(\beta)=\Omega_{\mathrm{eq}}-\Delta\Omega\sin^2\beta,
\end{equation}

\begin{figure}[b!!]
    \resizebox{\hsize}{!}
            {\includegraphics[clip]{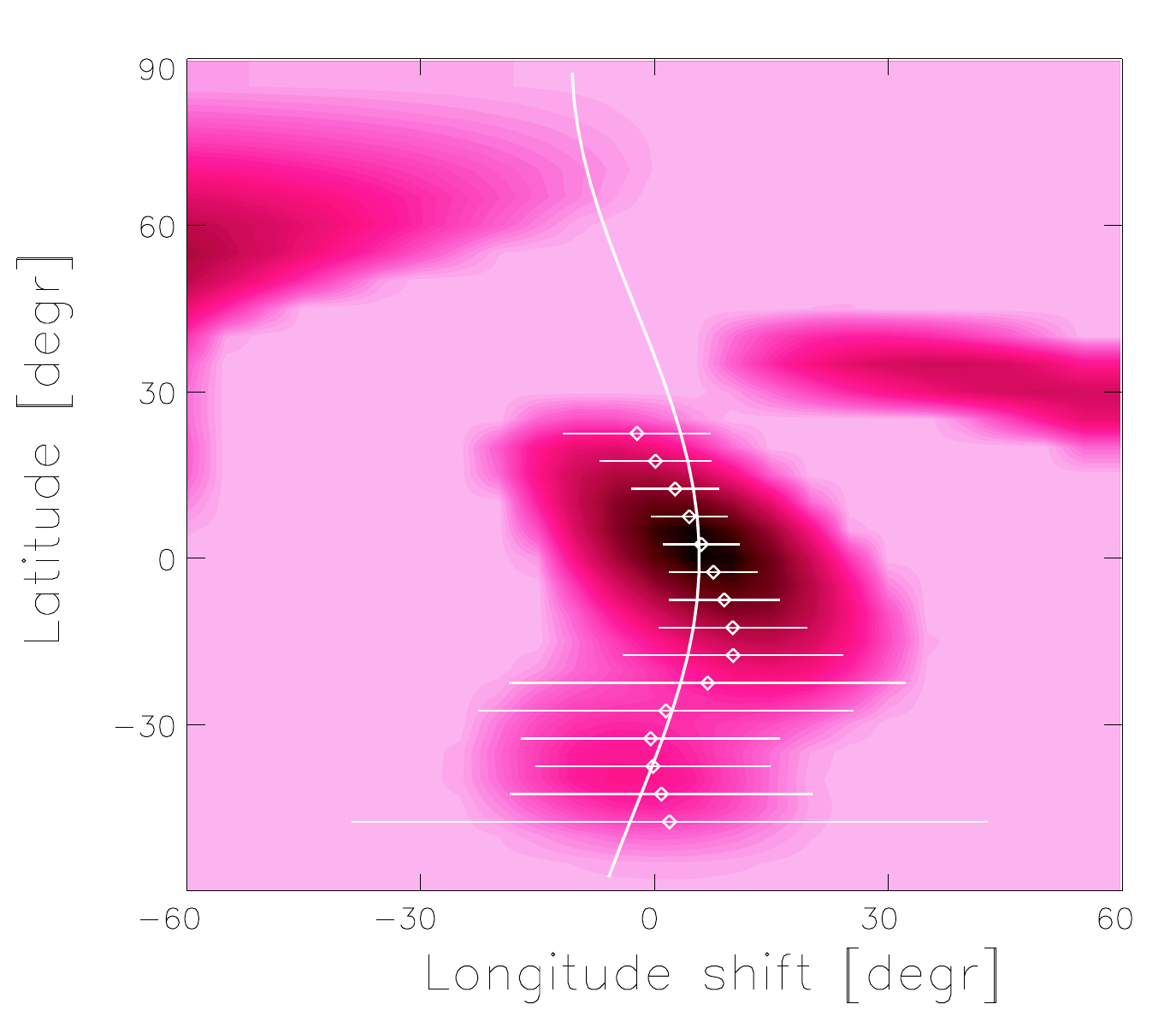}}
      \caption{Cross-correlation function map obtained by cross-correlating the two subsequent surface maps shown in Fig.~\ref{fig_di}. Darker regions correspond to stronger correlation. The best fit rotation law to the correlation peaks (dots) suggests solar-type differential rotation with a surface shear of $\alpha=0.016$.  
              }
         \label{fig_ccf}
\end{figure}

where $\Omega(\beta)$ is the angular velocity at $\beta$ latitude, $\Omega_{\mathrm{eq}}$ is the angular velocity of the equator, and $\Delta\Omega=\Omega_{\mathrm{eq}}-\Omega_{\mathrm{pole}}$ gives the difference between the equatorial and polar angular velocities. With these, the dimensionless surface shear parameter $\alpha$ is defined as $\alpha=\Delta\Omega/\Omega_{\mathrm{eq}}$.

We cross-correlate the available two Doppler images to build up a 2D cross-correlation function map shown in Fig.~\ref{fig_ccf}. 
Despite some noise it is clear from the figure that the equator rotates most rapidly and the rotation velocity decreases with increasing latitude angle, i.e., the correlation pattern indicates solar-type surface differential rotation on V1358\,Ori. When fitting the pattern with a solar-type differential rotation law of the quadratic form above, we get $\Omega_{\mathrm{eq}}=266.8\pm0.3\,{}^{\circ}/\mathrm{day}$ and $\Delta\Omega=4.3\pm1.0 \,{}^{\circ}/\mathrm{day}$ resulting in $\alpha=0.016\pm0.004$ surface shear parameter. Errors are estimated from the FWHMs and amplitudes of the Gaussian fits to the latitudinal bins. 
However, having only two consecutive Doppler images could introduce false correlation, making the cross-correlation technique less powerful \citep{kovari_diffrot}. Moreover, due to projection effect, the actual impact of the strongest polar features is restricted. Also, the low latitude features are less contrasted. As a result, the true errors of the derived surface shear may be somewhat larger, therefore we assume $\pm$0.01 error bar for $\alpha$ instead of the formal value of $\pm$0.004 from the fit.
For further discussion on the differential rotation, see Sect.~\ref{sect_disc}.

\section{Chromospheric activity}\label{sect_chrom}

We measured the Ca {\sc II} $R_{\mathrm{HK}}$, H$\alpha$ and Ca {\sc ii} IRT  chromospheric activity indices for all of the spectra individually to see if there is any rotational modulation present for the two covered rotations.

For the $R_{\mathrm{HK}}$, we first calculated the non-calibrated $S$-index as described in \cite{vaughan_s}. The instrumental values were then transformed into the original Mt. Wilson scale with the calibration coefficients for NARVAL derived by \cite{marsden_sindex}. To avoid color-dependence, $S$-index were transformed to $R_{\mathrm{HK}}$ (\citealt{middelkoop_rhk}, \citealt{rutten_rhk}). To subtract the photospheric adjunct, we applied the correction formula of \cite{noyes_rhk}. For the H$\alpha$, we used the indicator defined by \cite{kurster_halpha}. The IRT index was calculated using the formula of \cite{marsden_sindex}.

\begin{figure}[t!!]
    \resizebox{\hsize}{!}
            {\includegraphics[clip]{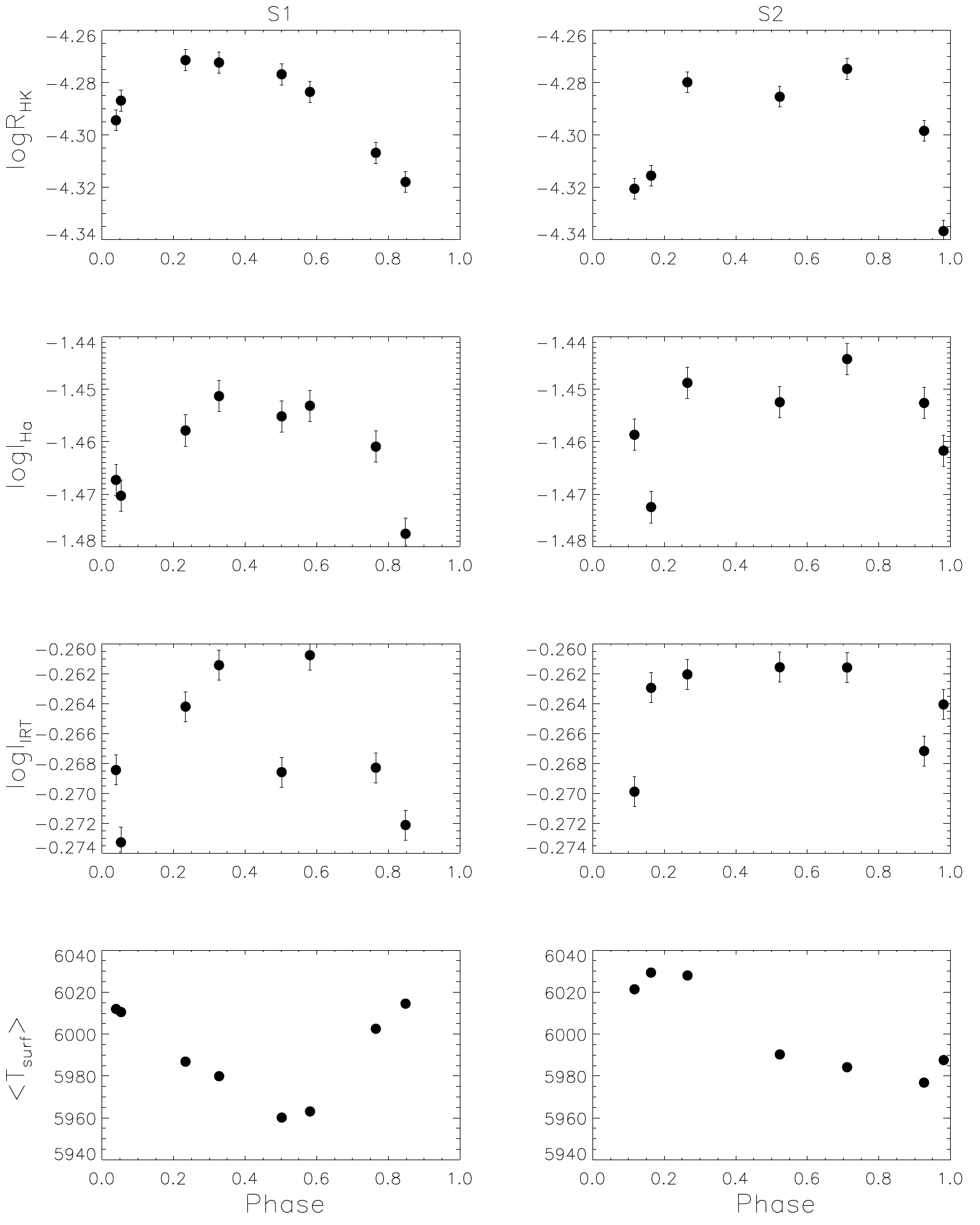}}
      \caption{ Calcium $R_{\mathrm{HK}}$ (top panels), H$\alpha$ (second row) and Ca {\sc ii} IRT (third row) curves for the two rotations compared to the average projected surface temperatures for the the same rotational phases from Doppler imaging (bottom panels).
              }
         \label{fig_chrom}
\end{figure}

The apparent average surface temperatures were also computed for the two Doppler images in the same phases for comparison. The $R_{\mathrm{HK}}$, H$\alpha$ and the IRT indices are plotted in Fig. \ref{fig_chrom} along with average temperatures. The values itself are summarized in Table \ref{table_chrom}. The errors were estimated using error propagation, and are in the order of 0.004, 0.003 and 0.001 for $R_{\mathrm{HK}}$, $\mathrm{H\alpha}$ and IRT, respectively.

All of the indices clearly show some change with the rotational phases which could be interpreted as rotational modulation. Apart from a few outlying points, the $R_{\mathrm{HK}}$ H$\alpha$ and IRT indices show roughly the same behaviour. The overall shape of the curves are similar, however, there is no clear correlation between the position of the maximal chromospheric activity and the highest spot coverage (i.e., the lowest average surface temperature). One might argue that the largest difference between the chromospheric activity and the average surface temperature is around 0.2--0.4 phase, where, on the Doppler images, the strongest hot spot becomes gradually visible, which might indicate that the hot structure has a chromospheric counterpart (and further strengthen the suspicion that these features are indeed not artifacts of the Doppler imaging process).

\section{Discussion}\label{sect_disc}

Both visual inspection and Fourier analysis suggest a possible activity cycle of $\approx\!1600\,\mathrm{d}$, i.e., $\approx\!4.5 \,\mathrm{yr}$. The additional Str\"omgren $y$ photometry (see Fig. \ref{fig_phot}) also seems to confirm this cycle, but these data were not included in the Fourier analysis, as the Str\"omgren $y$ filter is much narrower than the Johnson $V$ (23 vs 88\,nm) and there is also a large gap in those observations.  
The rotational period and the length of the activity cycle are two important observables of magnetically active stars. Their ratio is related to the dynamo number which is an indicator of magnetic activity. The derived rotational period and the long-term cycles (including the suspected $\approx 5200$ days long one) fits well to the rotational period -- activity cycle length relation derived by \cite{olah_pcyc} (see Fig. 5 and details therein). 

In subplot 3) ($\mathrm{HJD}=2452533-2452473$), 4) (2452894--2453111) and 8) (2454370--2454557) of Fig. \ref{fig_phot_all}, the amplitudes of the light curves are significantly lower than in the other seasons. This may be a prelude to the actual flip-flop phenomenon: the stronger active region gradually weakens, while the weaker one becomes stronger.

The suspected flip-flop period of $\approx\! 6\,\mathrm{years}$ would fit the  flip-flop periods found on single dwarfs by \cite{elstner_flipflop}. They pointed out that the presence of the flip-flop mechanism shows little to none dependence on the thickness of the convective layer (and hence the spectral type), however, they suspected a stronger dependence on the strength of the surface differential rotation. They found that the reported flip-flop periods of several years ($\approx\!4-9$) are found in the range of $\lvert\alpha\rvert=\lvert\Delta\Omega/\Omega_{\mathrm{eq}}\rvert\approx0.015-0.15$ (see \citealt{elstner_flipflop} and references therein). Our values ($\alpha=0.016\pm0.01$ and $P_{\mathrm{ff}}\approx6\,\mathrm{yr}$) fit this suspected relation well.

We note however, that between the last two plotted seasons, a season was omitted due to the low number of points, therefore the above period is an upper estimate.  

The surface of V1358\,Ori is dominated by a large polar structure of high contrast accompanied by a low number of low latitude, weaker features.  While polar features are not uncommon on active stars in general, this configuration seems to be a recurring phenomenon on young, single solar analogues. \cite{jarvinen_v889her} reported a very similar, and rather stable surface structure on V889\,Her, a young G2V star using Doppler imaging, and later \cite{kovari_v889her} derived a similar surface structure. \cite{marsden_v557car} also detected a dominant polar cap without any significant low-latitude features on V557\,Car, a roughly 40 Myr old G2 solar analogue. EK\,Dra, another young G2V star was also reported to exhibit these features (although here with some stronger low latitude spots as well) by \cite{jarvinen_ekdra}. \cite{hackman1} also derived strong polar structures and weaker equatorial features (and magnetic field configurations consistent with this picture) for V1358\,Ori and another two young solar-like stars, AH\,Lep and HD\,29615, (G3V and G2V respectively) using Zeeman-Doppler imaging.  It should be noted however, that on the maps of \cite{hackman1}, there were no bright features at all around the equator, which may indicate that the magnetic configuration has undergone structural changes.
\begin{table*}
\caption{Chromospheric activity indices of V1358 Ori in the observed rotational phases. See Sect. \ref{sect_chrom} for more details.      }     
\label{table_chrom}      
\centering                         
\begin{tabular}{c c c c || c c c c}        
\hline\hline
& S1  & & & & S2  & &\\
\hline\hline
$\phi$ & $\log{R_{\mathrm{HK}}}$ & $\log{I_{\mathrm{H\alpha}}}$ & $\log{I_{\mathrm{IRT}}}$ & $\phi$ & $\log{\mathrm{R}_{\mathrm{HK}}}$ & $\log{I_{\mathrm{H\alpha}}}$ & $\log{I_{\mathrm{IRT}}}$ \\
\hline
    0.039 &      -4.294 &      -1.467 &     -0.268 &      0.117 &      -4.320 &      -1.459 &     -0.270 \\
    0.053 &      -4.287 &      -1.470 &     -0.273 &      0.163 &      -4.315 &      -1.472 &     -0.263 \\
    0.233 &      -4.271 &      -1.458 &     -0.264 &      0.265 &      -4.280 &      -1.449 &     -0.262 \\
    0.327 &      -4.272 &      -1.451 &     -0.261 &      0.523 &      -4.285 &      -1.452 &     -0.262 \\
    0.502 &      -4.277 &      -1.455 &     -0.269 &      0.711 &      -4.275 &      -1.444 &     -0.262 \\
    0.581 &      -4.284 &      -1.453 &     -0.261 &      0.926 &      -4.298 &      -1.453 &     -0.267 \\
    0.765 &      -4.307 &      -1.461 &     -0.268 &      0.981 &      -4.337 &      -1.462 &     -0.264 \\
    0.848 &      -4.318 &      -1.478 &     -0.272 &  & & & \\

\hline
\end{tabular}
\end{table*}
Recently, \cite{isik_mod} used numerical simulations of magnetic flux transport and emergence to model the difference in latitudinal spot distributions between the Sun and stars. They found that as the rotation rate increases, magnetic flux emerges at higher latitudes, and a quiet region opens around the equator. They also pointed out that at 8 times of the rotational rate of the Sun ($P_{\mathrm{rot}}\!\approx\!3\,\mathrm{d}$) polar regions can form, while the width of the inactive region around the equator increases. Our findings are also consistent with this model.

Finally we emphasize that our result of differential rotation fits well to the observation that active young solar-type stars exhibit weak solar-type differential rotation, e.g: LQ Hya $\alpha=0.0056$ with $P_{\mathrm{rot}}=1.597\,\mathrm{d}$, \citealt{kovari_lqhya}) or AB Dor ($\alpha\approx0.006$ with $P_{\mathrm{rot}}=0.5148\,\mathrm{d}$), \citealt{jeffers_abdor}).
Moreover, \cite{kovari_v889her} derived $\alpha=0.009$ for V889\,Her, with a rotational period of 1.337 days and \cite{marsden_v557car} reported a surface shear of 0.012 on V557\,Car ($P_{\mathrm{rot}}=0.557\,\mathrm{d}$). Our result is also in good agreement with the empirical relation between the rotational period and the surface shear parameter of 
\begin{equation}
    \lvert\alpha\rvert \approx 0.005\,P_{\mathrm{rot}} [\mathrm{days}]
\end{equation}
for single stars suggested by \cite{kovari_diffrot}, see Fig. \ref{fig_diffrot}.

The rotational modulation of the chromospheric activity indicators suggest that the positions of the chromospheric structures (plages?) more-or-less coincide with the positions of the photospheric nests. The most apparent difference in the shape of the curves is in the region of 0.2--0.4 phase, which coincides with the phase where the most prominent hot feature on the Doppler images of both rotation becomes visible do to the rotation of the star. This may mean that the photospheric hot features extend to the chromosphere as well, and contribute to the overall chromospheric emission.

\section{Summary}
\begin{itemize}
    \item Based on a 14 years long photometric dataset we derive a rotational period of $P_{\mathrm{rot}}=1.3571\,\mathrm{d}$ for V1358\,Ori.
    \item An activity cycle with the period of roughly 1600 days is detected, which is consistent with the findings of \cite{olah_pcyc}. A flip-flop time scale of 6 years may also be present.
    \item By a spectral synthesis technique we determine precise astrophysical parameters for V1358\,Ori.
    \item We perform Doppler imaging to map the surface temperature distribution for two subsequent epochs separated by two weeks. The surface structure is dominated by a large polar cap accompanied with weaker features at low latitudes, consistent with previous observations of young solar analogues and recent dynamo models as well. Hot features are also present on both maps.
    \item Surface differential rotational is derived by cross-correlating the two subsequent Doppler images. The resulting surface shear parameter $\alpha=0.016\pm0.01$ fits to the rotational period-surface shear empirical relationship proposed recently by \cite{kovari_diffrot}.
    \item Chromospheric activity indicators are calculated and compared to the average apparent surface temperatures. Rotational modulation is present on the activity indicator curves in both rotations. The shapes of the curves are similar. The most prominent difference between the activity indicator curves and the average surface temperatures may indicate that the hot spots contribute to the chromospheric emission.
\end{itemize}

\begin{acknowledgements}
The authors acknowledge the Hungarian National Research, Development and Innovation Office
grant OTKA K-113117, and supports through
the Lend\"ulet-2012 Program (LP2012-31) of the Hungarian Academy of Sciences,
and the ESA PECS Contract No. 4000110889/14/NL/NDe. 
KV is supported by the Bolyai J\'anos Research Scholarship of the Hungarian Academy of Sciences. The authors thank A. Mo\'or for the useful conversations on stellar ages.  Finally, the authors would like to thank the anonymous referee for her/his valuable insights.
\end{acknowledgements}

%-------------------------------------------------------------------

\bibliography{bib}

\begin{appendix}
\section{}
\begin{figure*}[htt!!]
            \centering
            {\includegraphics[width=0.74\textwidth]{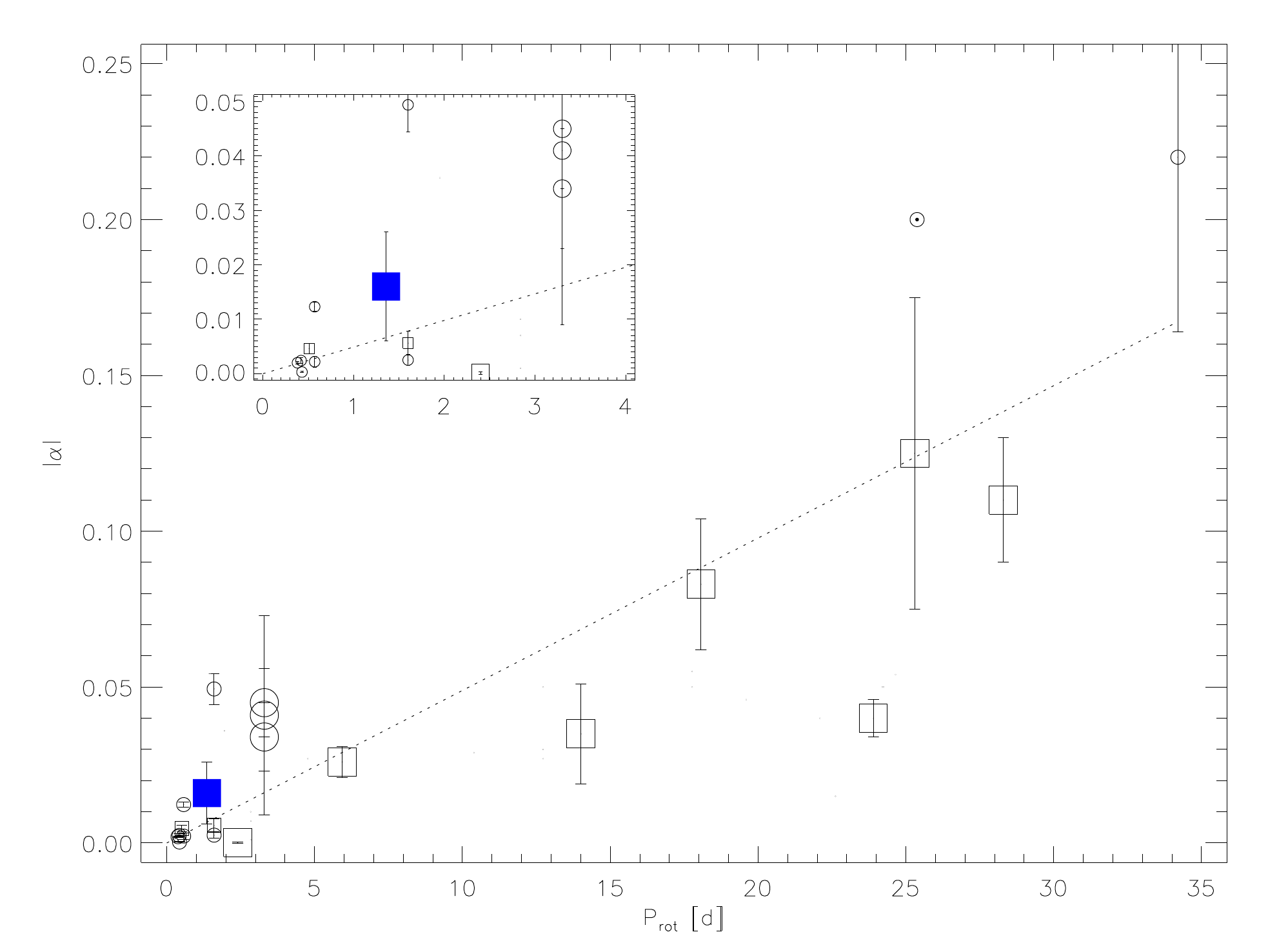}}
      \caption{The absolute value of the dimensionless surface shear parameter of  single stars plotted against their rotational period in days (see \citealt{kovari_diffrot} and references therein). Circles denote results from the sheared image method, while squares indicate values obtained with the cross-correlation technique. The blue filled square indicates V1358 Ori. The dotted line represents a linear fit with a steepness of $\approx0.005$. See Sect. \ref{sect_disc} for further details.
              }
         \label{fig_diffrot}
\end{figure*}
\begin{figure}
\resizebox{\hsize}{!}
            {\includegraphics[clip]{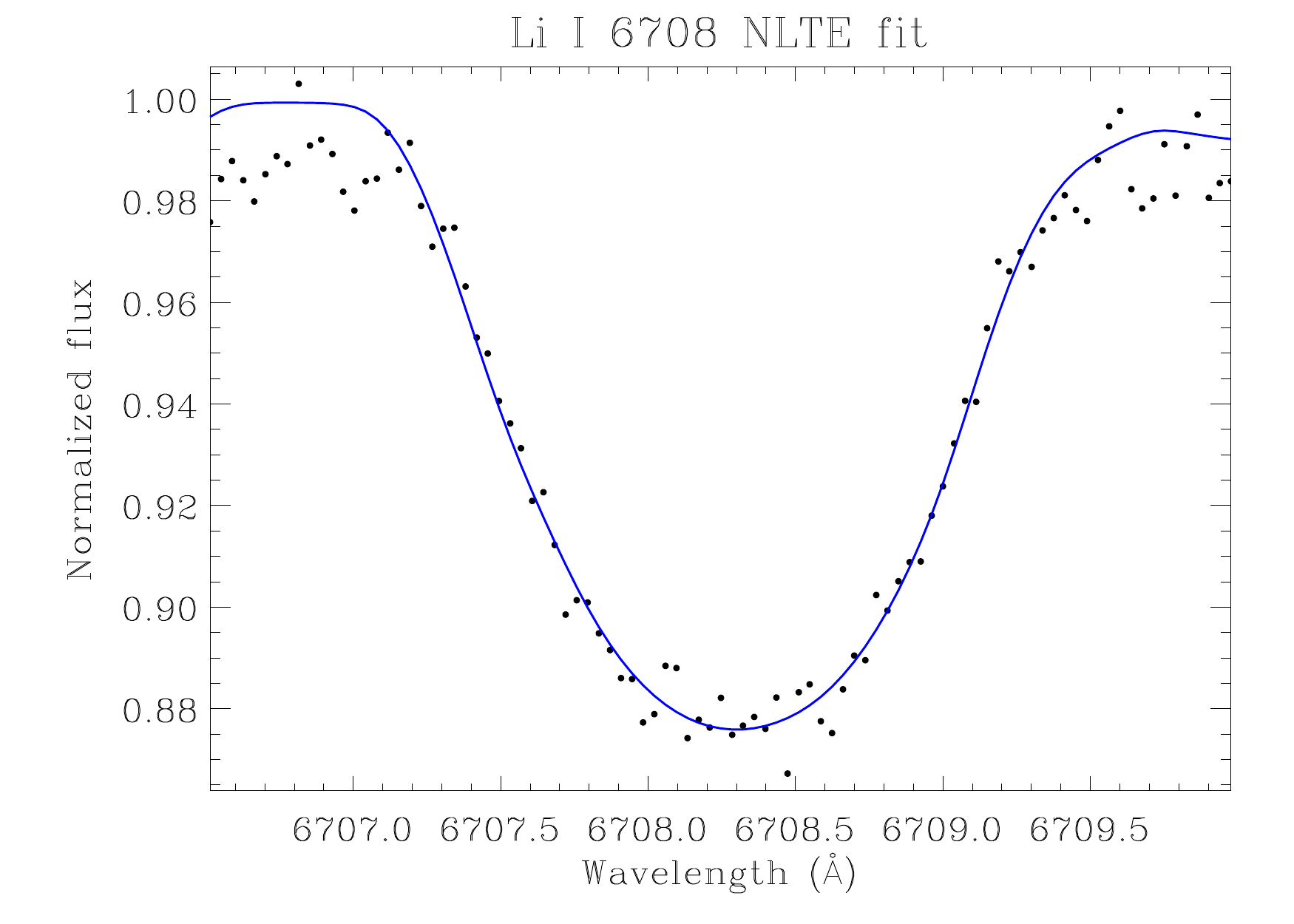}}
      \caption{An example NLTE Li {\sc i} 6708 fit. The corresponding lithium abundance is $A_{\mathrm{NLTE}}(\mathrm{Li})=2.17\pm0.03$. 
              }
         \label{fig_li}
\end{figure}

\end{appendix}

\end{document}